\documentclass{acm_proc_article-sp}
\usepackage[table]{xcolor}
\usepackage{tikz}
\usetikzlibrary{shapes.geometric}
\usepackage{arydshln}
\usepackage{listings}

\lstdefinelanguage{ACME}
{
 morecomment = [l]{//}, 
 morecomment = [l]{///},
 morecomment = [s]{/*}{*/},
 morestring=[b]", 
 sensitive = true,
 keywords = {and,Attachments,boolean,Component,Connector,Dependencies,Detach,do,extends,extended,false,Family,Forall,from,in,Invariant,new,On,Port,Property,Remove,Role,System,true,to,Type,with}
}
\lstdefinelanguage{PDDL}{}
\begin{document}


\title{ACME vs PDDL: support for dynamic reconfiguration of software architectures}

\numberofauthors{3}
\author{
  \alignauthor Jean-Eudes M\'ehus\\
  \affaddr{\'Ecoles de St-Cyr Co\"etquidan}\\
  \affaddr{Guer, France}\\
  \email{jean-eudes.mehus@st-cyr.terre-net.defense.gouv.fr}
  \alignauthor Thais Batista\\
  \affaddr{Federal University of Rio Grande do Norte}\\
  \affaddr{Natal (RN) Brazil}\\
  \email{thais@dimap.ufrn.br}
  \alignauthor J\'er\'emy Buisson\\
  \affaddr{UEB / \'Ecoles de St-Cyr Co\"etquidan / Universit\'e de Bretagne Sud}\\
  \affaddr{Guer, France}\\
  \email{jeremy.buisson@st-cyr.terre-net.defense.gouv.fr}
}

\maketitle

\begin{abstract}
On the one hand, ACME is a language designed in the late 90s as an
interchange format for software architectures. The need for
reconfiguration at runtime has led to extend the language with
specific support in Plastik. On the other hand, PDDL is a predicative
language for the description of planning problems. It has been
designed in the AI community for the International Planning
Competition of the ICAPS conferences. Several related works have
already proposed to encode software architectures into PDDL. Existing
planning algorithms can then be used in order to generate
automatically a plan that updates an architecture to another one,
i.e., the program of a reconfiguration. In this paper, we improve the
encoding in PDDL. Noticeably we propose how to encode ADL types and
constraints in the PDDL representation. That way, we can statically
check our design and express PDDL constraints in order to ensure that
the generated plan never goes through any bad or inconsistent
architecture, not even temporarily.
\end{abstract}

\category{D.2}{Software}{Software Engineering}
\keywords{dynamic reconfiguration, ACME, PDDL, planning, software architecture}

\newcommand{\todo}[1]{\textcolor{red}{[TODO: #1]}}

\section{Introduction}

With long-running software systems, we sometimes want to change the
software program at runtime. That way, one can deploy new features or
fix bugs without any service disruption. Continuous operation and
critical systems have such requirements for dynamic
reconfiguration. Dynamic reconfiguration have been studied at
different levels (control flow~\cite{Duquesne+Bryce:Stage},
functions~\cite{Armstrong:Erlang,Neamtiu+EtAl:Ginseng},
objects~\cite{2003:Appavoo+EtAl:K42,Dmitriev:JavaHotSwap},
components~\cite{Plastik,Fractal,FScript}). In this paper we consider
only structural changes of the component-based software
architecture. At this level, dynamic reconfigurations typically
consist in adding and removing components and connectors, as well as
changing the connections between architectural elements.

Several software architecture reconfiguration languages have been
propo\-sed~\cite{Plastik,FScript,CommUnity} to let developers program
reconfigurations by-hand. These languages provide primitive operations
that add, remove and modify architectural elements at runtime.  The
developer uses these operations in combination with usual connectives
(sequence, iteration, condition) to describe imperatively how to
transform the software architecture to the new desired version.

In order to relieve from the burden of programming reconfigurations,
some related works have tried to automate the issuing of
reconfiguration
instructions~\cite{Daubert,Arshad,ArshadSQJ,ElMaghraoui,PDDLDeployment,AOMDA}. These
techniques compare the original and target architectures in order to
identify the changes. Then a tool computes a suitable sequence of
primitive operations that performs the reconfiguration.

Instead of designing domain-specific algorithms, it is appealing to
reuse off-the-shelf techniques to generate reconfiguration
scripts. That way we can benefit from advances and expertise in other
research fields. Automatic action planning from the AI community is
one candidate technology. It is a field of research that focuses on
the generation of a sequence of actions that brings a system from an
initial state to a goal state, based on the specification of all
possible actions. Planners such as POPF~\cite{POPF}, Fast
Downward~\cite{FastDownward}, LAMA~\cite{LAMA} and
Madagascar~\cite{Madagascar} are general-purpose tools that can be
used if software architecture reconfigurations can be translated to
planning problems.

In the AI community, the \emph{de facto} standard description language
for planning problems is PDDL~\cite{pddl}. This language is designed
and used by the International Planning Competition of the ICAPS
conferences. PDDL is therefore a widely spread language that is
implemented by many planners.

As noticed by Andr\'e \emph{et al.}~\cite{Daubert}, PDDL let us switch
easily from one planner to another. Furthermore, regular planners
usually generate better reconfiguration scripts than simple
domain-specific heuristics such as~\cite{AOMDA}. However, in the
current state of the art, constraints (coming from either the software
architecture, the component model or the execution platform) are not
taken into account. Even if the specification of reconfiguration
operations is correctly designed, there is no guaranty (in the current
state of the art) that the planner cannot generate operations that
infringe any constraint. In order to address this issue, we propose in
this paper that we statically verify general constraints and that we
embed the other constraints in the planning problem.

\begin{table*}
  \centering
  \caption{ACME extensions for programmed reconfiguration.}
  \begin{tabular}{|l|p{9cm}|}\hline
    Reconfiguration statements & Description \\ \hline
    \lstinline+On (<Armani exp>) do {<statements>}+ & expresses runtime conditions under which programmed reconfigurations should take place, and a specification (in terms of the other reconfiguration statements) of what should change. \\ \hline
    \lstinline+Detach <element> from <element>+ & removes an attachment between a port and a role. \\ \hline
    \lstinline+Remove <element>+ & destroys an existing component, connector or representation. \\ \hline
    \lstinline+Dependencies {<ACME statements>}+ & expresses runtime dependencies among components/connectors (e.g., if X is to be removed, Y should be removed also). \\ \hline
  \end{tabular}
  \label{acme:reconf}
\end{table*}

This paper improves the current state of the art (described in
Section~\ref{state-of-the-art}) in order to take into account
constraints. Our presentation is based on a synthetic client-server
example described in Section~\ref{example}. We program this example
using the ACME~\cite{acme} architecture description language in
Section~\ref{acme}. As exposed, ACME is richer than the architecture
description languages used by Arshad \emph{et al.}, Andr\'e \emph{et
  al.} and Ingstrup \emph{et al}. Indeed, ACME supports the
component-and-connector paradigm with types, as well as invariants and
architectural styles. By means of its extension Armani~\cite{Armani},
ACME let the software architect state constraints. ACME has also
built-in support for dynamic reconfiguration thanks to
Plastik~\cite{Plastik}. In this paper, we do not refrain from defining
new reconfiguration operations, e.g., modifying
types. Section~\ref{pddl} describes the same example using PDDL. We
first presents this language. We give the specification of the
primitive reconfiguration operations as well as the predicates that we
use to encode software architectures. A set of invariants relates the
predicates according to the ACME component model. In
Section~\ref{invariants} we show how some of the constraints coming
either from the ACME language itself (e.g., typing) or from
architectural styles (e.g., client-server) can be checked statically,
and therefore we show that the planning problem is consistent with
these invariants. The remaining invariants can be embedded in the
planning problem in order to tell the planner not to infringe any of
the constraints. Section~\ref{conclusion} contains a discussion of our
results, reports our first experiments and concludes the paper with
future works.

\section{State of the art}
\label{state-of-the-art}

In the current state of the art, regular PDDL planners from the AI
community have already been used successfully to generate
automatically reconfiguration
scripts~\cite{Daubert,Arshad,ArshadSQJ,ElMaghraoui,PDDLDeployment}. With
these systems, the original architecture and the target one are
encoded together as a planning problem. Each reconfiguration primitive
is modelled as an action in a planning domain. From these
descriptions, a planner generates a sequence (possibly partially
ordered if concurrency is supported) of primitive operations that
brings the architecture from its original state to the target
configuration. It is important that the planning problem states
clearly the constraints imposed by, e.g., the execution platform in
order that the planner does not generate an infeasible plan.

We build on the previous state-of-the-art results of Arshad \emph{et
  al.}~\cite{Arshad,ArshadSQJ}, Andr\'e \emph{et al.}~\cite{Daubert},
Ingstrup \emph{et al.}~\cite{PDDLDeployment} and El Maghraoui \emph{et
  al.}~\cite{ElMaghraoui} in order to improve their techniques. All of
these previous works translate reconfigurations to PDDL planning
problems. They define a set of predicates to describe a software
architecture as the conjunction of logical facts. For example
in~\cite{ArshadSQJ}, the \lstinline+connected-component+ predicate
states whether a given component and a given connector are connected
to one another. A set of actions models the semantics (preconditions
and effects) of the reconfiguration primitive operations. An action
named \lstinline+connect-component+ for instance requires that a
component and a connector are instantiated and that none of them is
connected; its effect establishes the \lstinline+connected-component+
fact. Andr\'e \emph{et al.}, Ingstrup \emph{et al.} and El Maghraoui
\emph{et al.}~\cite{ElMaghraoui} do the same even if they define
different sets of predicates and actions, taking into account their
respective contexts (i.e., their component model and their execution
platform).

El Maghraoui \emph{et al.}~\cite{ElMaghraoui} propose in addition an
encoding for properties that is based on predicates. A predicate named
\lstinline!set! states whether a given property of a given object has
a value. For each property of each type, a specific predicates relates
an object to the value of the attribute.

In summary, the main achievements and limits of previous works are:
\begin{itemize}
\item Arshad \emph{et al.}~\cite{Arshad,ArshadSQJ} use the approach in
  the context of a component-and-connector ADL. However, their model
  does not take into account ports, roles, types, constraints or
  architectural styles.
\item Andr\'e \emph{et al.}~\cite{Daubert} use the approach in the
  context of DiVA ART~\cite{DiVA:ART}. They support component
  instances and types as well as ports. However, they do not have the
  connector concept and they do not support constraints or
  architectural styles.
\item Ingstrup \emph{et al.}~\cite{PDDLDeployment} have done similar
  experiments on the generation of deployment plans for OSGi
  bundles. However, they do not consider types or architectural
  styles.
\item El Maghraoui \emph{et al.}~\cite{ElMaghraoui} use PDDL planning
  in order to generate deployment plans for datacenters managed by
  tools such as IBM Tivoli Provisioning Manager~\cite{Tivoli}. They
  support object-relationship models with properties. While types are
  taken into account, their structure is not modelled in PDDL. They do
  not support constraints.
\end{itemize}

Each of these previous works assume consistency rules that relate the
predicates. For instance, El Maghraoui \emph{et
  al.}~\cite{ElMaghraoui} assume the consistency rules that states
that if a property is set for an object then the property has a value
for that object; at there is at most one value per property-and-object
pair; and so on. Only few of such rules are explicitly given. While
Ingstrup \emph{et al.} have analyzed their reconfiguration actions
using Alloy, they acknowledge that the version used with AI planning
is different from the verified one~\cite{PDDLAlloy}. None of Arshad
\emph{et al.}~\cite{Arshad,ArshadSQJ}, Andr\'e \emph{et
  al.}~\cite{Daubert} or El Maghraoui \emph{et al.}~\cite{ElMaghraoui}
take any preventive measure in regard to such constraints. None of
these works avoid that the planner generates inconsistent, ill-typed
or non-conformant architectures as intermediate reconfiguration
steps. This is the point we address in this paper.

\section{A running example}
\label{example}

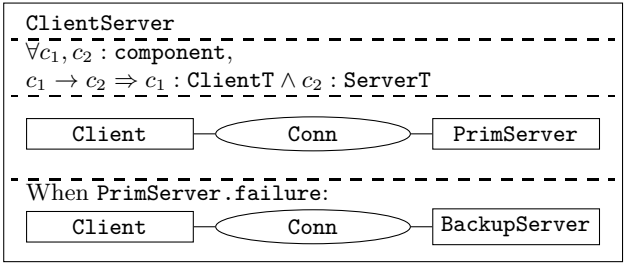
\begin{figure}
  \centering
  \begin{tikzpicture}
    \node[draw,rectangle] at (0,0) {
      \begin{tabular}{l}
        \lstinline!ClientServer!\\
        \hdashline
        $\forall c_1, c_2 : \textrm{\lstinline!component!},$\\$c_1 \rightarrow c_2 \Rightarrow c_1: \textrm{\lstinline!ClientT!} \wedge c_2: \textrm{\lstinline!ServerT!}$\\
        \hdashline
        \begin{tikzpicture}[baseline=(current bounding box.north)]
          \node[align=center,draw,text width=2cm] (Client) at (0,0) {\lstinline!Client!};
          \node[align=center,draw,ellipse,text width=1.7cm,inner sep=.2em] (Conn) at (2.7,0) {\lstinline!Conn!};
          \node[align=center,draw,text width=2cm] (PrimServer) at (5.4,0) {\lstinline!PrimServer!};
          \draw (Client) -- (Conn) -- (PrimServer);
        \end{tikzpicture}\\ \\
        \hdashline
        When \lstinline!PrimServer.failure!:\\
        \begin{tikzpicture}
          \node[align=center,draw,text width=2cm] (Client) at (0,0) {\lstinline!Client!};
          \node[align=center,draw,ellipse,text width=1.7cm,inner sep=.2em] (Conn) at (2.7,0) {\lstinline!Conn!};
          \node[align=center,draw,text width=2cm] (BackupServer) at (5.4,0) {\lstinline!BackupServer!};
          \draw (Client) -- (Conn) -- (BackupServer);
        \end{tikzpicture}      \end{tabular}
    };
  \end{tikzpicture}
  \caption{Architecture of the running example.}
  \label{fig:example}
\end{figure}

Figure~\ref{fig:example} depicts the architecture that we use as the
running example of this paper. It is a client-server architecture,
which contains two components (\lstinline+Client+ is the client;
\lstinline+PrimServer+ is the server) and a connector named
\lstinline+Conn+. An invariant is attached to this architecture in
order to ensure the client-server style. It states that for any pair
$c_1, c_2$ of components, if $c_1$ is connected to $c_2$, then $c_1$
conforms to type \lstinline+ClientT+ and $c_2$ conforms to type
\lstinline+ServerT+. This invariant prevents from connecting one
client component to another client, or one server component to another
server.

Regarding reconfiguration, one may want to replace the primary server
component \lstinline+PrimServer+ in case it fails. Obviously the
reconfigured architecture, i.e., using a backup server named
\lstinline!BackupServer! in place of \lstinline+PrimServer+, conforms
to the invariant.

In PDDL, the states of a system, including the initial state and the
goal state, are described by formulas in the first-order predicate
logic. A planner may generate any plan that conforms to the
goal. Conformance is defined by implication: as long as the final
state implies the goal, the plan is acceptable. Consequently, the
resulting architecture is not exactly the same as the goal
architecture. For example, the resulting architecture may contain
components, connectors and bindings that are not required, as long as
they are not forbiden by the goal formula.

Therefore if we don't take any preventive measure, despite the target
architecture respects invariants and constraints, the resulting
architecture (after effective execution of the generated
reconfiguration) may not.

A planner may execute any action as long as it can be executed. For
instance, a planner may freely trigger an action, then undo that
action at the next step. While planner implementors do their best not
to issue useless actions, it is not mandatory that a planner generate
optimal plans.

Therefore if we don't take any preventive measure, some intermediate
steps shall be invalid with respect to invariants and constraints even
if the resulting architecture is conformant.

In our example, we would like that the connector is detached from
\lstinline!PrimServer! before it is connected to the backup server;
that the backup server is instantiated before its port is bound. The
reconfiguration designer may also want to state that the architecture
must continuously conform to its invariant during the reconfiguration;
or a specific invariant shall be given for the time of the
reconfiguration.

While the current state-of-the-art
techniques~\cite{Daubert,Arshad,ArshadSQJ,ElMaghraoui,PDDLDeployment}
propose carefully designed planning problems, none of these previous
works describe how to verify that they enforce the
constraints. Despite they verify reconfiguration operations using
Alloy~\cite{PDDLAlloy}, Ingstrup \emph{et al.} acknowledge that the
PDDL specification is not the verified one. In addition, in case some
constraints cannot be checked statically, none of these previous works
propose how to embed the constraints in the planning problem. None
of~\cite{Daubert,Arshad,ArshadSQJ,ElMaghraoui,PDDLAlloy,PDDLDeployment}
take into account invariants given by the software architect.

\section{The example using ACME ADL}
\label{acme}

ACME~\cite{acme}  is an extensible generic ADL that provides a syntax for representing structures and an annotation mechanism for describing additional semantics. The ACME core concepts are: 
\begin{itemize}
\item Components: the basic building blocks in an ACME description of a system. Components expose their functionality through their ports. A port represents a point of contact between the component and its environment.
\item Connectors: represent communication glue that captures the nature of an interaction between components. \emph{Ports} are bound to ports on other components using connectors. Like components, connectors may be used to model a variety of different sorts of interactions under a number of different models. A connector includes a set of interfaces in the form of  \emph{roles}.
\item Systems: describe a set of components and connectors, and how they interact. A property might also be used to represent properties of the environment in which the system is operating, or ``global'' properties that apply to all elements of the system. The graph of a system (how everything is connected) is defined by a set of \emph{attachments}.
\item Representations: are alternative decompositions of a given component; they reify the notion that a component may have multiple alternative implementations. Elements in ACME may have more than one representation.
\item Element types: are defined in the same way as instances; they define a prototype that is instantiated by copying its structure. The ACME type model states that all instances of a type must include the structure defined by this type. For properties, this means that if a set of properties is defined for a particular type, any instances must have the same properties.  
\item Attachments: define a set of port/role associations. 
\item  Properties: are <name, type, value> triples that annotate to any of the above ACME elements.  ACME allows user-defined property types that may be defined in terms of built-in property data types. Systems, components, connectors, ports and roles may include a list of properties and a list of representations.
\item Architectural styles: define sets of types of components, connectors, properties, and sets of rules that specify how elements of those types may be legally composed in a reusable architectural domain.
\end{itemize}
Armani~\cite{Armani} extends ACME with a language based on first-order pre\-dicates used to express architectural constraints over architectures. For example, it can be used to express constraints on system composition, behaviour, and properties. Constraints are defined in terms of so-called invariants which in turn are composed of standard logical connectives and predicates (both built-in and user-defined) which are referred to as functions.

ACME may also be used as a way of representing reconfigurable architectures by expressing the possible reconfigurations in terms of the ACME structures. For example, a system might include properties that describe components that may be added at run-time and how to attach them to the current system. This means that ACME does not include first class elements to describe dynamic reconfiguration of the architecture. In other words, dynamic reconfiguration is not originally addressed by ACME but it can be handled using the extensible mechanism of the language. 
In order to address this issue, Plastik~\cite{Plastik} defines ACME extensions to represent different types of reconfigurations at the architecture level. Table~\ref{acme:reconf} summarizes the set of such ACME extensions for programmed reconfiguration, that is, reconfiguration that can be foreseen at design time. Ad-hoc reconfiguration, which involves changes unforeseen at design time, can be specified at the architecture level by submitting a partial architecture specification to a configurator. At the architecture level Plastik is based on styles to constrain the allowable range of permissible ad-hoc reconfigurations. 

Figure~\ref{plastik} illustrates the use of Armani and Plastik in our example. A programmed reconfiguration specifies the removal of the \lstinline!PrimServer! component if it fails: the \lstinline!Conn! connector and the \lstinline!PrimServer! component are disconnected, then \lstinline!PrimServer! is removed, and a \lstinline!BackupServer! component is inserted and attached to \lstinline!Conn!. 

\begin{figure}
\scriptsize\begin{lstlisting}[fontadjust,basewidth=.45em]
Family ClientServerFam extends PlastikMF with {
 Component Type ClientT = { Port request = new RequiredPort; }
 Component Type ServerT = { Port service = new ProvidedPort; }
 Invariant Forall c1, c2: Component in self.Components |
  connected(c1, c2) -> (satisfiesType(c1, ?ClientT?) and
                        satisfiesType(c2, ?ServerT?));
}
System ClientServer = new ClientServerFam extended with {
 Component Client = new ClientT;
 Component PrimServer = new ServerT extended with {
  Property failure: boolean = false;
 }
 Connector Conn = { Role requestor; Role servicer; }
 Attachments { Client.request to Conn.requestor;
               Conn.servicer to PrimServer.service; }
 On (PrimServer.failure == true) do {
  Detach Conn.servicer from PrimServer.service;
  Remove PrimServer;
  Component BackupServer = new ServerT extended with {
   Dependencies {
    Attachments { Conn.servicer to BackupServer.service; }
  }
 }
}
\end{lstlisting}\normalsize
\caption{A dependable client-server in Plastik.}
\label{plastik}
\end{figure}

The description of an architecture in ACME with dynamic reconfiguration statements follows the following steps:
\begin{itemize}
\item Identify the concepts in the source model that correspond to the ACME architectural concept: system, component, connector, port, role or representation
\item Define a family or set of families for the model. Define a component, connector, port or role type to represent each of the architectural concepts
\item Define a set of property types, which will make up a property language for describing elements in the model
\item Define the reconfiguration actions  
\end{itemize}

\section{The example using PDDL}
\label{pddl}

PDDL is a standard encoding language for planning tasks.  The
components of a PDDL planning task are: (i) \emph{Objects} (ii)
\emph{Predicates}: properties of objects that can be true or false, as
in first order logic; (iii) \emph{Initial state}: the list of all
facts that are true in the initial state; (iv) \emph{Goal
  specification}: the objective of the problem that specifies what
need to be true at the end of the plan; (v) \emph{State trajectory
  constraints}: a logical formula used to restrict the space of
states; (vi) \emph{Actions/Operators}: ways of changing the truth and
falsity of facts. Actions are parameterized with objects. An action is
composed of a \emph{precondition}, that states the constraints to the
action be executed, and an \emph{effect} that lists the facts that
become true or false after the execution of the action.

The planning task is usually split into two files:
\begin{enumerate}
\item A domain file for the definition of the domain predicates and
  actions. To some extent, the domain defines the language used to
  describe situations and planning problems in a specific application.
\item A problem file containing the objects of the problem instance,
  initial state and goal specifications.
\end{enumerate}

In summary, the description of predicates structures the
representation of states; the description of actions characterizes
domain behaviours. Predicates and actions (the domain) are separated
from the description of specific instance objects, initial conditions,
and goals that characterize a problem instance. A planning problem is
created by joining a domain description with a problem
description. The same domain description can be joined with many
different problem descriptions to yield different planning problems in
the same domain. The structure of a domain file is depicted in
Figure~\ref{domain}. The structure of a problem file is depicted in
Figure~\ref{problem}.

\begin{figure}
\scriptsize\begin{lstlisting}[language=PDDL,fontadjust,basewidth=.45em]

(define (domain <domain name>)
   <PDDL code for predicates>
   <PDDL code for first action>
   [...]
   <PDDL code for last action>
)
\end{lstlisting}\normalsize
\caption{Domain File in PDDL.}
\label{domain}
\end{figure}

\begin{figure}
\scriptsize\begin{lstlisting}[language=PDDL,fontadjust,basewidth=.45em]
(define (problem <problem name>)
   (:domain <domain name>)
   <PDDL code for objects>
   <PDDL code for initial state>
   <PDDL code for goal specification>
)
\end{lstlisting}\normalsize
\caption{Problem File in PDDL.}
\label{problem}
\end{figure}

\begin{table*}
  \centering
  \caption{Predicates for containment relationships.}
  \begin{tabular}{|r|p{3cm}|p{3cm}|p{3cm}|p{3cm}|} \cline{2-5}
    \multicolumn{1}{c}{} & \multicolumn{2}{|c|}{Types} & \multicolumn{2}{|c|}{Instances} \\ \hline
    system & \multicolumn{2}{|c|}{\cellcolor{black!20}} & \multicolumn{1}{|c|}{\lstinline!constains-component!} & \multicolumn{1}{|c|}{\lstinline!contains-connector!} \\ \hline
    component/connector & \multicolumn{1}{|c|}{\lstinline!type-has-port!} & \multicolumn{1}{|c|}{\lstinline!type-has-role!} & \multicolumn{1}{|c|}{\lstinline!has-port!} & \multicolumn{1}{|c|}{\lstinline!has-role!} \\ \hline
  \end{tabular}
  \label{tab:containment}
\end{table*}

In this work, we inspire from previous works: most architecture
elements (components, connectors, ports, roles, systems and types) are
encoded as objects. The relationships between elements (including
attachments) are encoded as facts, thanks to predicates. Each
reconfiguration primitive operation reflects as a PDDL
action. Architectural styles are either checked statically or
programmed as trajectory constraints. In this paper, we ignore
representations and properties. We do not consider the reconfiguration
of invariants or styles.

\subsection{A domain for software architectures}
\label{txt:domain}

Like~\cite{ArshadSQJ}, we use PDDL types only to classify objects
depending on the kind of architecture elements. The reason why we
don't encode ACME types using PDDL types is because the two type
systems are different. On the one side, ACME component and connector
types are structural types: a component is an instance of a type if
and only if it contains at least the same structure as the one
described in the type. On the other side, PDDL types are nominal
types. The type of an object is given by name. Two types with
different names are different types.

The relationship between components and connectors instances on the
one side, and types on the other side is modelled by
\lstinline!has-component-type! and \lstinline!has-connector-type!.

The containment relationships are implemented by a set of predicates,
one per level in the hierarchy. Table~\ref{tab:containment} summarizes
these predicates. Our model assumes that if a component \lstinline!c!
has type \lstinline!t!, then \lstinline!c! contains the same ports as
\lstinline!t!. The same invariant applies to connectors and roles.

The \lstinline!bound! predicate binds a port of a component to a role
of a connector. The predicate named \lstinline!exist-component!
(resp. \lstinline!exist-connector!) states that a component
(resp. connector) is instantiated.

We also define negative predicates: the predicate named
\lstinline!unbound-port! (resp. \lstinline!unbound-role!) states that
a port of a component (resp. a role of a connector) is not bound to
any role (resp. port).

\newcommand{\defop}[4]{\item \expandafter\lstinline\expandafter!#1!: #2.\\Precondition: #3.\\Effects: #4.}
We model each reconfiguration primitive operation as an action,
expliciting the preconditions and effects on the architecture in terms
of the above predicates.
\begin{itemize}
\defop{create-component}{instantiate a component named \lstinline!?c! in a system \lstinline!?s!}{$\neg\textrm{\lstinline!exist-component!}\left(\textrm{\lstinline!?c!}\right)$}{$\textrm{\lstinline!exist-component!}\left(\textrm{\lstinline!?c!}\right)$,\\ $\textrm{\lstinline!contains-component!}\left(\textrm{\lstinline!?s!},\textrm{\lstinline!?c!}\right)$}
\defop{create-connector}{instantiate a connector named \lstinline!?c! in a system \lstinline!?s!}{$\neg\textrm{\lstinline!exist-connector!}\left(\textrm{\lstinline!?c!}\right)$}{$\textrm{\lstinline!exist-connector!}\left(\textrm{\lstinline!?c!}\right)$,\\ $\textrm{\lstinline!contains-connector!}\left(\textrm{\lstinline!?s!},\textrm{\lstinline!?c!}\right)$}
\defop{remove-component}{remove a component named \lstinline!?c! from a system \lstinline!?s!}{$\textrm{\lstinline!contains-component!}\left(\textrm{\lstinline!?s!},\textrm{\lstinline!?c!}\right)$}{$\neg\textrm{\lstinline!exist-component!}\left(\textrm{\lstinline!?c!}\right)$,\\ $\neg\textrm{\lstinline!contains-component!}\left(\textrm{\lstinline!?s!},\textrm{\lstinline!?c!}\right)$}
\defop{remove-connector}{remove a connector \lstinline!?c! from a system \lstinline!?s!}{$\textrm{\lstinline!contains-connector!}\left(\textrm{\lstinline!?s!},\textrm{\lstinline!?c!}\right)$}{$\neg\textrm{\lstinline!exist-connector!}\left(\textrm{\lstinline!?c!}\right)$,\\ $\neg\textrm{\lstinline!contains-connector!}\left(\textrm{\lstinline!?s!},\textrm{\lstinline!?c!}\right)$}
\defop{attach}{bind a port \lstinline!?p! of a component \lstinline!?c! to a role \lstinline!?r! of a connector \lstinline!?co!}{%
  $\textrm{\lstinline!exist-component!}\left(\textrm{\lstinline!?c!}\right)$,\\
  $\textrm{\lstinline!exist-connector!}\left(\textrm{\lstinline!?co!}\right)$,
  $\textrm{\lstinline!has-port!}\left(\textrm{\lstinline!?c!},\textrm{\lstinline!?p!}\right)$,\\
  $\textrm{\lstinline!has-role!}\left(\textrm{\lstinline!?co!},\textrm{\lstinline!?r!}\right)$,
  $\textrm{\lstinline!unbound-port!}\left(\textrm{\lstinline!?c!},\textrm{\lstinline!?p!}\right)$,\\
  $\textrm{\lstinline!unbound-role!}\left(\textrm{\lstinline!?co!},\textrm{\lstinline!?r!}\right)$}{%
  $\neg\textrm{\lstinline!unbound-port!}\left(\textrm{\lstinline!?c!},\textrm{\lstinline!?p!}\right)$,\\
  $\neg\textrm{\lstinline!unbound-role!}\left(\textrm{\lstinline!?co!},\textrm{\lstinline!?r!}\right)$,
  $\textrm{\lstinline!bound!}\left(\textrm{\lstinline!?c!},\textrm{\lstinline!?p!},\textrm{\lstinline!?co!},\textrm{\lstinline!?r!}\right)$}
\defop{detach}{unbind a port \lstinline!?p! of a component \lstinline!?c! from a role \lstinline!?r! of a connector \lstinline!?co!}{%
  $\textrm{\lstinline!bound!}\left(\textrm{\lstinline!?c!},\textrm{\lstinline!?p!},\textrm{\lstinline!?co!},\textrm{\lstinline!?r!}\right)$}{%
  $\neg\textrm{\lstinline!bound!}\left(\textrm{\lstinline!?c!},\textrm{\lstinline!?p!},\textrm{\lstinline!?co!},\textrm{\lstinline!?r!}\right)$,\\
  $\textrm{\lstinline!unbound-port!}\left(\textrm{\lstinline!?c!},\textrm{\lstinline!?p!}\right)$,
  $\textrm{\lstinline!unbound-role!}\left(\textrm{\lstinline!?co!},\textrm{\lstinline!?r!}\right)$}
\end{itemize}

As we model the types of architectural elements, we can also define
operations that affect types. As in ACME each element has its own type
(defined by its own structure) we can define operations that
consistently change a component or a connector and its type at
once. No question arises whether such modifications should propagate
to a whole group of instances as each type is bound to one instance at
most.
\begin{itemize}
\defop{add-port}{add a port \lstinline!?p! to a component \lstinline!?c! of type \lstinline!?t!}{%
  $\textrm{\lstinline!has-component-type!}\left(\textrm{\lstinline!?c!},\textrm{\lstinline!?t!}\right)$,\\
  $\neg\textrm{\lstinline!has-port!}\left(\textrm{\lstinline!?c!},\textrm{\lstinline!?p!}\right)$}{%
  $\textrm{\lstinline!has-port!}\left(\textrm{\lstinline!?c!},\textrm{\lstinline!?p!}\right)$,
  $\textrm{\lstinline!type-has-port!}\left(\textrm{\lstinline!?t!},\textrm{\lstinline!?p!}\right)$,\\
  $\textrm{\lstinline!unbound-port!}\left(\textrm{\lstinline!?c!},\textrm{\lstinline!?p!}\right)$}
\defop{add-role}{add a role \lstinline!?r! to a connector \lstinline!?c! of type \lstinline!?t!}{%
  $\textrm{\lstinline!has-connector-type!}\left(\textrm{\lstinline!?c!},\textrm{\lstinline!?t!}\right)$,\\
  $\neg\textrm{\lstinline!has-role!}\left(\textrm{\lstinline!?c!},\textrm{\lstinline!?r!}\right)$}{%
  $\textrm{\lstinline!has-role!}\left(\textrm{\lstinline!?c!},\textrm{\lstinline!?r!}\right)$,
  $\textrm{\lstinline!type-has-role!}\left(\textrm{\lstinline!?t!},\textrm{\lstinline!?r!}\right)$,\\
  $\textrm{\lstinline!unbound-role!}\left(\textrm{\lstinline!?c!},\textrm{\lstinline!?r!}\right)$}
\defop{remove-port}{remove a port \lstinline!?p! from a component \lstinline!?c! of type \lstinline!?t!}{%
  $\textrm{\lstinline!has-component-type!}\left(\textrm{\lstinline!?c!},\textrm{\lstinline!?t!}\right)$,\\
  $\textrm{\lstinline!has-port!}\left(\textrm{\lstinline!?c!},\textrm{\lstinline!?p!}\right)$,
  $\textrm{\lstinline!unbound-port!}\left(\textrm{\lstinline!?c!},\textrm{\lstinline!?p!}\right)$}{%
  $\neg\textrm{\lstinline!has-port!}\left(\textrm{\lstinline!?c!},\textrm{\lstinline!?p!}\right)$,\\
  $\neg\textrm{\lstinline!type-has-port!}\left(\textrm{\lstinline!?t!},\textrm{\lstinline!?p!}\right)$}
\defop{remove-role}{remove a role \lstinline!?r! from a connector \lstinline!?c! of type \lstinline!?t!}{%
  $\textrm{\lstinline!has-connector-type!}\left(\textrm{\lstinline!?c!},\textrm{\lstinline!?t!}\right)$,\\
  $\textrm{\lstinline!has-role!}\left(\textrm{\lstinline!?c!},\textrm{\lstinline!?r!}\right)$,
  $\textrm{\lstinline!unbound-role!}\left(\textrm{\lstinline!?c!},\textrm{\lstinline!?r!}\right)$}{%
  $\neg\textrm{\lstinline!has-role!}\left(\textrm{\lstinline!?c!},\textrm{\lstinline!?r!}\right)$,\\
  $\neg\textrm{\lstinline!type-has-role!}\left(\textrm{\lstinline!?t!},\textrm{\lstinline!?r!}\right)$}
\end{itemize}

In case a type is not bound to any instance, we can also reconfigure
that type independently of any instance.
\begin{itemize}
\defop{add-type-port}{add a port \lstinline!?p! to a component type \lstinline!?t!}{\\%
  $\forall\textrm{\lstinline!?c!}:\textrm{\lstinline!component!},\neg\textrm{\lstinline!has-component-type!}\left(\textrm{\lstinline!?c!},\textrm{\lstinline!?t!}\right)$,\\
  $\neg\textrm{\lstinline!type-has-port!}\left(\textrm{\lstinline!?t!},\textrm{\lstinline!?p!}\right)$}{%
  $\textrm{\lstinline!type-has-port!}\left(\textrm{\lstinline!?t!},\textrm{\lstinline!?p!}\right)$}
\defop{add-type-role}{add a role \lstinline!?r! to a connector type \lstinline!?t!}{\\%
  $\forall\textrm{\lstinline!?c!}:\textrm{\lstinline!connector!},\neg\textrm{\lstinline!has-connector-type!}\left(\textrm{\lstinline!?c!},\textrm{\lstinline!?t!}\right)$,\\
  $\neg\textrm{\lstinline!type-has-role!}\left(\textrm{\lstinline!?t!},\textrm{\lstinline!?r!}\right)$}{%
  $\textrm{\lstinline!type-has-role!}\left(\textrm{\lstinline!?t!},\textrm{\lstinline!?r!}\right)$}
\defop{remove-type-port}{remove a port \lstinline!?p! from a component type \lstinline!?t!}{\\%
  $\forall\textrm{\lstinline!?c!}:\textrm{\lstinline!component!},\neg\textrm{\lstinline!has-component-type!}\left(\textrm{\lstinline!?c!},\textrm{\lstinline!?t!}\right)$,\\
  $\textrm{\lstinline!type-has-port!}\left(\textrm{\lstinline!?t!},\textrm{\lstinline!?p!}\right)$}{%
  $\neg\textrm{\lstinline!type-has-port!}\left(\textrm{\lstinline!?t!},\textrm{\lstinline!?p!}\right)$}
\defop{remove-type-role}{remove a role \lstinline!?r! from a connector type \lstinline!?t!}{\\%
  $\forall\textrm{\lstinline!?c!}:\textrm{\lstinline!connector!},\neg\textrm{\lstinline!has-connector-type!}\left(\textrm{\lstinline!?c!},\textrm{\lstinline!?t!}\right)$,\\
  $\textrm{\lstinline!type-has-role!}\left(\textrm{\lstinline!?t!},\textrm{\lstinline!?r!}\right)$}{%
  $\neg\textrm{\lstinline!type-has-role!}\left(\textrm{\lstinline!?t!},\textrm{\lstinline!?r!}\right)$}
\end{itemize}

It is the role of the constraints such as subtyping or type
satisfaction to restrict the use of the operation that affect types.

In this design, we assume that any situation conforms to the following
consistency constraints. The equality $=$ denotes the identity over
PDDL objects (including components, connectors, their types, ...).
\begin{itemize}
\item Each component has a type:
  \[
  \begin{array}{l}
    \forall c: \textrm{\lstinline!component!}, \exists t:
    \textrm{\lstinline!component-type!},\\
    \textrm{\lstinline!has-component-type!}\left(c, t\right)
  \end{array}
  \]
\item The type of a component is unique:
  \[
  \begin{array}{l}
    \forall c: \textrm{\lstinline!component!}, \forall t_1, t_2:
    \textrm{\lstinline!component-type!},\\
    \textrm{\lstinline!has-component-type!}\left(c, t_1\right)\\
    \wedge \textrm{\lstinline!has-component-type!}\left(c, t_2\right)\\
    \Rightarrow t_1 = t_2
  \end{array}
  \]
\item Each component has a different type:
  \[
  \begin{array}{l}
    \forall c_1, c_1: \textrm{\lstinline!component!}, \forall t:
    \textrm{\lstinline!component-type!},\\
    \textrm{\lstinline!has-component-type!}\left(c_1, t\right)\\
    \wedge \textrm{\lstinline!has-component-type!}\left(c_2, t\right)\\
    \Rightarrow c_1 = c_2
  \end{array}
  \]
\item A component and its type have the same ports:
  \[
  \begin{array}{l}
    \forall c: \textrm{\lstinline!component!}, \forall t:
    \textrm{\lstinline!component-type!},\\
    \textrm{\lstinline!has-component-type!}\left(c, t\right) \\
    \Rightarrow
    \forall p: \textrm{\lstinline!port!},\\
    \textrm{\lstinline!has-port!}\left(c, p\right) \Leftrightarrow \textrm{\lstinline!type-has-port!}\left(t, p\right)
  \end{array}
  \]
\item A system contains only instantiated components:
  \[
  \begin{array}{l}
    \forall s: \textrm{\lstinline!system!},
    \forall c: \textrm{\lstinline!component!},\\
    \textrm{\lstinline!contains-component!}\left(s, c\right)\\
    \Rightarrow \textrm{\lstinline!exist-component!}\left(c\right)
  \end{array}
  \]
\item Any instantiated component lies in a system:
  \[
  \begin{array}{l}
    \forall c: \textrm{\lstinline!component!},
    \textrm{\lstinline!exist-component!}\left(c\right) \\
    \Rightarrow
    \exists s: \textrm{\lstinline!system!},
    \textrm{\lstinline!contains-component!}\left(s, c\right)
  \end{array}
  \]
\item A component is in a single system:
  \[
  \begin{array}{l}
    \forall c: \textrm{\lstinline!component!},
    \forall s_1, s_2: \textrm{\lstinline!system!},\\
    \textrm{\lstinline!contains-component!}\left(s_1, c\right)\\
    \wedge \textrm{\lstinline!contains-component!}\left(s_2, c\right)\\
    \Rightarrow s_1 = s_2
  \end{array}
  \]
\item Only instantiated components can be bound:
  \[
  \begin{array}{l}
    \forall c: \textrm{\lstinline!component!},
    \forall p: \textrm{\lstinline!port!},
    \forall co: \textrm{\lstinline!connector!},\\
    \forall r: \textrm{\lstinline!role!},
    \textrm{\lstinline!bound!}\left(c, p, co, r\right)
    \Rightarrow \textrm{\lstinline!exist-component!}\left(c\right)
  \end{array}
  \]
\item Only the ports of a component can be bound:
  \[
  \begin{array}{l}
    \forall c: \textrm{\lstinline!component!},
    \forall p: \textrm{\lstinline!port!},
    \forall co: \textrm{\lstinline!connector!},\\
    \forall r: \textrm{\lstinline!role!},
    \textrm{\lstinline!bound!}\left(c, p, co, r\right)
    \Rightarrow \textrm{\lstinline!has-port!}\left(c, p\right)
  \end{array}
  \]
\item A port of a component can be bound only once:
  \[
  \begin{array}{l}
    \forall c: \textrm{\lstinline!component!},
    \forall p: \textrm{\lstinline!port!},\\
    \forall co_1, co_2: \textrm{\lstinline!connector!},
    \forall r_1, r_2: \textrm{\lstinline!role!},\\
    \textrm{\lstinline!bound!}\left(c, p, co_1, r_1\right)
    \wedge \textrm{\lstinline!bound!}\left(c, p, co_2, r_2\right)\\
    \Rightarrow co_1 = co_2 \wedge r_1 = r_2
  \end{array}
  \]
\item A port is unbound iff it is not bound to any role:
  \[
  \begin{array}{l}
    \forall c: \textrm{\lstinline!component!},
    \forall p: \textrm{\lstinline!port!},\\
    \left(
    \begin{array}{l}
      \textrm{\lstinline!unbound-port!}\left(c, p\right) \\
      \Leftrightarrow
      \forall co: \textrm{\lstinline!connector!},
      \forall r: \textrm{\lstinline!role!},\\
      \neg\textrm{\lstinline!bound!}\left(c, p, co, r\right)    
    \end{array}
    \right)
  \end{array}
  \]
\end{itemize}

\begin{figure}
\scriptsize\begin{lstlisting}[language=PDDL,fontadjust,basewidth=.45em]
(:objects ClientServer             - system
          Client PrimServer        - component
          BackupServer             - component
          Conn                     - connector
          ClientT Client-type      - component-type
          ServerT PrimServer-type  - component-type
          BackupServer-type        - component-type
          Conn-type                - connector-type
          request service          - port
          requestor servicer       - role)
\end{lstlisting}\normalsize
  \caption{The client-server PDDL objects.}
  \label{fig:cs:objects:pddl}
\end{figure}

The same constraints hold for connectors and roles.

\subsection{The architecture in our PDDL domain}

Figure~\ref{fig:cs:objects:pddl} gives the PDDL listing for the objects
in the architecture. It enumerates all of the architectural elements
that are mapped onto PDDL objects: systems, components, connectors,
ports, roles, component types and connector types. It contains all the
objects that could exist before, during and after the reconfiguration.

Figure~\ref{fig:cs:init:pddl} defines the client-server of our running
example using our PDDL domain. It extensively states the facts that
are true in the architecture. The facts that are not listed are
assumed false. This conjonction of facts conforms to the constraints
that we have identified in the PDDL domain.

\begin{figure}
\scriptsize\begin{lstlisting}[language=PDDL,fontadjust,basewidth=.45em]
(:init (exist-component Client)
       (exist-component PrimServer)
       (exist-connector Conn)
       (contains-component ClientServer Client)
       (contains-component ClientServer PrimServer)
       (contains-connector ClientServer Conn)
       (has-component-type Client Client-type)
       (has-component-type PrimServer PrimServer-type)
       (has-connector-type Conn Conn-type)
       (has-port Client request)
       (has-port PrimServer service)
       (has-role Conn requestor)
       (has-role Conn servicer)
       (type-has-port Client-type request)
       (type-has-port PrimServer-type service)
       (type-has-role Conn-type requestor)
       (type-has-role Conn-type servicer)
       (type-has-port ClientT request)
       (type-has-port PrimServer-type service)
       (bound Client request Conn requestor)
       (bound PrimServer service Conn servicer)
       (has-component-type BackupServer BackupServer-type))
\end{lstlisting}\normalsize
  \caption{The client-server architecture in PDDL.}
  \label{fig:cs:init:pddl}
\end{figure}

\begin{figure}
\scriptsize\begin{lstlisting}[language=PDDL,fontadjust,basewidth=.45em]
(:goal (and (exist-component Client)
            (exist-component BackupServer)
            (exist-connector Conn)
            (contains-component ClientServer Client)
            (contains-component ClientServer BackupServer)
            (contains-connector ClientServer Conn)
            (has-component-type Client Client-type)
            (has-component-type BackupServer BackupServer-type)
            (has-connector-type Conn Conn-type)
            (has-port Client request)
            (has-port BackupServer service)
            (has-role Conn requestor)
            (has-role Conn servicer)
            (type-has-port Client-type request)
            (type-has-port BackupServer-type service)
            (type-has-role Conn-type requestor)
            (type-has-role Conn-type servicer)
            (type-has-port ClientT request)
            (type-has-port BackupServer-type service)
            (bound Client request Conn requestor)
            (bound BackupServer service Conn servicer)))
\end{lstlisting}\normalsize
  \caption{The reconfigured client-server architecture in PDDL.}
  \label{fig:cs:goal:pddl}
\end{figure}

Following the same principle, Figure~\ref{fig:cs:goal:pddl} defines the
client-server of our running example after the reconfiguration. This
is the goal clause of the PDDL problem file. With this example, the
generated plan is the one given in Figure~\ref{fig:cs:plan:pddl}:
\begin{enumerate}
\item First, a \lstinline!service! port is added to the
  \lstinline!BackupServer! component. Indeed this port is absent in
  the initial architecture of Figure~\ref{fig:cs:init:pddl}.
\item Second, the \lstinline!service! port of the
  \lstinline!PrimServer! component is detached from the
  \lstinline!servicer! role of the \lstinline!Conn! connector.
\item Third, the \lstinline!BackupServer! component is instantiated
  within the \lstinline!ClientServer! system.
\item Last, the \lstinline!service! port of the
  \lstinline!BackupServer! component is bound to the
  \lstinline!servicer! role of the \lstinline!Conn! connector.
\end{enumerate}
As the goal does not state that the \lstinline!PrimServer! component
should be destroyed, the generated plan does not executes the
\lstinline!remove-component! action. Except few differences as
noticed, the generated plan is almost the same as the hand-written
reconfiguration of Figure~\ref{plastik}. According to the
specification of the actions, the \lstinline!add-port! and
\lstinline!create-component! actions can be executed in any arbitrary
order.

\begin{figure}
\scriptsize\begin{lstlisting}[language=PDDL,fontadjust,basewidth=.45em]
(add-port BackupServer BackupServer-type service)
(detach PrimServer service Conn servicer)
(create-component ClientServer BackupServer)
(attach BackupServer service Conn servicer)
\end{lstlisting}\normalsize
  \caption{The generated reconfiguration plan.}
  \label{fig:cs:plan:pddl}
\end{figure}

Architectural styles are translated as additional constraints in the
problem file.

\section{Checking invariants}
\label{invariants}

During a reconfiguration, the software system passes by a succession
of intermediate architectures, at the end of each primitive
reconfiguration operation. Several invariants must hold in any
architecture: some of them are inherent of the ADL itself; some of
them come from the architectural style. Even if we choose to relax
these constraints temporarily during reconfiguration, we have to
ensure that the still-required properties are always satisfied. For
instance, Fractal~\cite{Fractal} and Andr\'e \emph{et
  al.}~\cite{Daubert} forbid a component to be active if any of its
client ports is not bound. Depending on the implementation, some
orderings of the ACME \lstinline!Detach! / \lstinline!Attachments!
might be forbiden as well, e.g., to prevent a client from being
temporarily bound to two servers at the same time.

We consider two strategies in order to enforce invariants.

On the one side, we can encode the invariants as trajectory
constraints, which have been introduced in
PDDL3~\cite{constraints}. Constraints are used to prune the search
space of the planner. The underlying logic is equivalent to a limited
subset of LTL. Temporal operators can be used to constrain dynamic
architectures. As ACME does not support dynamic architectures,
temporal operators are not needed in order to encode invariants. For
instance, the clause in Figure~\ref{fig:client-server-pddl} gives the
PDDL syntax in order to encode the invariant for the client-server
style.
\begin{figure}
  \centering
\scriptsize\begin{lstlisting}[language=PDDL,fontadjust,basewidth=.45em]
(:constraints (always (and
 (forall (?c1 ?c2              - component
          ?co                  - connector
          ?request ?service    - port
          ?requestor ?servicer - role
          ?c1T ?c2T            - component-type
   (implies
     (and (has-port ?c1 ?request)
          (has-port ?c2 ?service)
          (has-role ?co ?requestor)
          (has-role ?co ?servicer)
          (bound ?c1 ?request ?co ?requestor)
          (bound ?c2 ?service ?co ?servicer)
          (has-component-type ?c1 ?c1T)
          (has-component-type ?c2 ?c2T))
     (and (forall (?p - port) (implies (type-has-port ClientT ?p)
                                       (type-has-port ?c1T ?p)))
          (forall (?p - port) (implies (type-has-port ServerT ?p)
                                       (type-has-port ?c2T ?p))))))))))
\end{lstlisting}\normalsize
  \caption{The client-server invariant of Figure~\ref{plastik} as a PDDL
    trajectory constraint.}
  \label{fig:client-server-pddl}
\end{figure}
It states that if any two components \lstinline!?c1! and
\lstinline!?c2! are connected by a connector \lstinline!?co!, then
their respective types are included in \lstinline!ClientT! and
\lstinline!ServerT!, respectively.

The Armani language for ACME invariants is based on the first-order
predicate logic, restricted to quantification over finite sets only,
like PDDL. The only restriction is therefore the ability to encode the
Armani primitive functions using our PDDL predicates. For instance, in
the list of primitive functions of~\cite{Armani}, the
\lstinline!satisfiesType! function can be implemented as the
verification that the element has the same subelements as declared in
the type. However, in our system, we do not provide the necessary
mechanism for the \lstinline!declaresType! function. Indeed, we do not
keep track of subtyping declarations in the PDDL encoding. We would
need to define additional predicates in order to implement the
\lstinline!declaresType! function.

On the other side, some of the invariants can be checked
statically. Indeed, if the primitive reconfiguration operations are
properly modelled, they should at least preserve the invariants coming
from the ADL, whatever the context. Therefore, we aim at proving that
primitive operations are consistent with the invariants. Let
$\mathcal{I}$ be the invariant. Given an operation with parameters
$p$, if the precondition $P\left(p\right)$ holds, then the invariant
must still be satisfied after the effect in
$\left[E\left(p\right)\right]\mathcal{I}$:
\[
\mathcal{I} \Rightarrow \forall p, P\left(p\right) \Rightarrow \left[E\left(p\right)\right]\mathcal{I}
\]

As a simple example, we may want to ensure that any port is bound to
at most one role and that no action can infringe that rule. This
property is formalized as:
\[
\begin{array}{l}
\forall c : \textrm{\lstinline!component!}, \forall p : \textrm{\lstinline!port!},\\ \forall r_1, r_2 : \textrm{\lstinline!role!}, \forall co_1, co_2 : \textrm{\lstinline!connector!}, \\
\textrm{\lstinline!bound!}\left(c, p, co_1, r_1\right) \wedge
\textrm{\lstinline!bound!}\left(c, p, co_2, r_2\right)\\ \Rightarrow co_1 = co_2 \wedge r_1 = r_2
\end{array}
\]
Any operation that does not have any \lstinline!bound! positive effect
obviously preserves this invariant. When we check the
\lstinline!attach!  operation, the positive effect
$\textrm{\lstinline!bound!}\left(\textrm{\lstinline!?c!},\textrm{\lstinline!?p!},\textrm{\lstinline!?co!},\textrm{\lstinline!?r!}\right)$
is established only if the precondition
$\textrm{\lstinline!unbound-port!}\left(\textrm{\lstinline!?c!},\textrm{\lstinline!?p!}\right)$
is true. The invariant holds as our predicates are such that $\forall
c, \forall p, \left(\textrm{\lstinline!unbound-port!}\left(c,p\right)
\Leftrightarrow \forall co, \forall r, \neg
\textrm{\lstinline!bound!}\left(c,p,co,r\right)\right)$.

Of course, the system has to be checked against this property and all
other constraints of section~\ref{txt:domain} as well.

Only invariants that are general, at the level of the architecture
description language, can be verified statically. Indeed, we involve
solely the PDDL domain, i.e., the specification of predicates and
reconfiguration actions, which is common to all of the possible
architectures.

\section{Conclusion}
\label{conclusion}

In this paper, we propose a schema to encode an ACME architecture
using the PDDL language. This work has several interests. First, as
pointed out by related works, the approach allows using automatic
planner from the AI community in order to generate automatically
reconfiguration scripts. Second, it lets us study reconfiguration
operations that do not exist currently in Plastik, e.g., operations
that affect the type of architectural elements. Third, we give a sound
semantics to our reconfiguration framework.

In comparison to related works, we improve the technique in that we
propose how to manage the type of architecture elements, as well as
architectural styles and constraints. We furthermore formally state
consistency constraints for our PDDL domain. We explain how all of
these constraints can either be checked statically or used to restrict
the state space of the planner.

\begin{figure}
\centering\scriptsize\begin{lstlisting}
(:derived (unbound-port ?c - component ?p - port)
  (forall (?co - connector ?r - role)
    (not (bound ?c ?p ?co ?r))))
(:derived (unbound-role ?co - connector ?r - role)
  (forall (?c - component ?p - port)
    (not (bound ?c ?p ?co ?r))))
\end{lstlisting}\normalsize
  \caption{PDDL definition of derived predicates.}
  \label{fig:derived}
\end{figure}

In this paper, we do several design choices:
\begin{itemize}
\item We decide not to use PDDL derived predicates, i.e., predicates
  that are given as a formula of the other predicates. We can use this
  feature for instance for the \lstinline!unbound-port! and
  \lstinline!unbound-role! predicates. Using the definitions of
  Figure~\ref{fig:derived}, we do not have to explicitly manage these
  predicates in the effects of \lstinline!attach! and
  \lstinline!detach!. We can implement the \lstinline!exist-component!
  and \lstinline!exist-connector! predicates as derived predicates
  using \lstinline!contains-component! and
  \lstinline!contains-connector! as well. However, while derived
  predicates are appealing, only few planners support this fragment of
  the PDDL language.
\item We translate straightforwardly the invariants of the software
  architecture into PDDL constraints. As a consequence, the invariants
  must hold during the whole reconfiguration plan; and no
  reconfiguration action can change the set of enforced invariants.
\item We assign a unique type to each component or connector. That
  way, we can safely elude the question: what happens to its type and
  to the components (resp. connectors) that share the same type when a
  port (resp. role) is added or removed to a component
  (resp. connector).
\item We ignore some fragments of the ACME architecture description
  language. We do not consider properties, representations, type
  satisfaction declarations, families. Representations and type
  satisfaction declarations are relations between architectural
  elements; families are types for systems. We can therefore use the
  same approach as for the other kinds of relations. The properties
  store values of primitive types (integer, floa\-ting-point number,
  string, boolean), enumerated types or constructed types (sequence,
  set, record). Storing values is supported by PDDL fluents, which are
  functions that map a value to PDDL objects. However, fluents are
  restricted to either objects (which may suit well, e.g., enumerated
  types) or numbers. Instead of fluents, we can use the schema of El
  Maghraoui \emph{et al.}~\cite{ElMaghraoui}.
\end{itemize}

This work is still in early stage. As short-term future work, we plan
to experiment our PDDL domain with real planners. Indeed, it is
well-known that almost no existing planner implements the whole PDDL
standard, as reported at the International Planning
Competition'2011. In our preliminary results using
$65$~planners\footnote{$55$~competing planners and $3$~non competing
  planners come from the International Planning Competition'2011
  public subversion repository; $7$~planners are downloaded from
  public web sites. Some of these $65$~planners differ only in the
  heuristics and parameters they use.}, only $17$~planners pass the
running example of this paper (without using constraints). Among them,
$14$~planners generate the $4$~actions plan of
Figure~\ref{fig:cs:plan:pddl}\footnote{Some planners generate a
  different sequence, but the actions are the same. Some planners
  propose to execute several of these actions in parallel. We have
  verified that all of the generated solutions are correct.}; the
$3$~other planners generate additional useless actions. Only
$1$~planner succeeds when we use derived predicates; no planner
supports constraints. Among the $48$~failing planners, $6$~planners
report that they do not support negation in action preconditions. We
still need to investigate the reason why the $42$~remaining planners
fail.

Even if more experiments confirm that no planner support all of the
PDDL features that we use, we notice that: negative preconditions can
be removed by the introduction of additional predicates and effects;
derived predicates can be expanded like we do in this paper, e.g., for
\lstinline!unbound-port!; quantifiers can be expanded as
quantification is over finite sets; constraints can be encoded into
preconditions. Furthermore, these features are not implemented
probably due to the fact that they are not used in the International
Planning Competition. We can therefore expect that they will be
supported as the competition evolves next years.

The performance of planners is also affected by some metrics in the
domain definition such as maximum number of positive or negative
effects and preconditions~\cite{STRIPS}. We therefore have to ensure
that existing planners behave correctly with our improvements. During
our first experiments, simple reconfigurations are solved in less than
$300$ms with an Intel E5400 / Linux x64 PC, except one planner that
takes $2.5$s. In our future work, we will evaluate the planners with
more complex reconfigurations.

\bibliographystyle{abbrv}
\bibliography{refs}

\end{document}